# Some new insights on the statistical origin of the Bekenstein-Hawking entropy
# Ⅰ: Some heuristic arguments

Zhi-Yong Wang[*]

*School of Optoelectronic Information, University of Electronic Science and Technology of China, Chengdu 610054, China*

Now that the Bekenstein–Hawking entropy has been found within the traditional theory, its physical interpretation should hide in quantum field theory in curved spacetime (only its quantum corrections require a detailed knowledge of quantum gravity). However, to interpret the statistical origin of the Bekenstein–Hawking entropy, there have developed many microscopic pictures, some of them are based on the theories that themselves remain to be experimentally verified, while some cite the Hawking temperature formula and actually argue in circle (and then they have not provided more information than the seminal work of Bekenstein and Hawking). Within quantum field theory in curved spacetime (without introducing any new hypothesis), it is inevitable for us to attribute the Bekenstein–Hawking entropy to entanglement entropy. In the first part we just provide a heuristic argument, a rigorous theoretical model will be presented in our next work.

PACS number (s): 04.70.Dy, 04.62.+v

## I. INTRODUCTION

Since the pioneer work of Bekenstein [1] and Hawking [2], we have understood that black holes behave as thermodynamic objects, have a temperature (Hawking temperature)

$$k_B T_H = \frac{\hbar c^3}{8\pi GM} = \frac{\hbar \kappa}{2\pi c}, \qquad (1)$$

and an entropy (Bekenstein–Hawking entropy)

$$S_{BH} = \frac{4\pi GM^2}{\hbar c} = \frac{A_h}{4l_p^2}, \qquad (2)$$

where $k_B$ is the Boltzmann constant, $G$ the gravitational constant, $c$ the velocity of light in vacuum, $\hbar$ the reduced Planck constant, $l_p = \sqrt{\hbar G/c^3}$ the Planck length in four dimensional (4D) spacetime, $\kappa = GM/r_s^2$ the surface gravity, $M$ and $A_h = 4\pi r_s^2$ are the mass and horizon area of the black hole ($r_s = 2GM/c^2$ is the Schwarzschild radius, for simplicity we take Schwarzschild black holes for example). *For the convenience of quantitative examination, we apply the international system of units*, but the temperature is measured in energy units, such that the factor $k_B$ is not included in the definition of entropy.

On the other hand, as we know, entropy is related to degrees of freedom and counts the number of microstates. However, classically, black holes possess very few degrees of freedom: the theorem that a black hole 'has no hair' tells us that, the only labels a black hole possesses are mass, angular momentum and electric charge, such that two Schwarzschild black holes with the same mass must be identical, and similarly two Kerr black holes with the same mass and angular momentum must be identical, etc [3]. So what is the statistical origin of black hole entropy? where are the degrees of freedom giving rise to black hole entropy located?

Until about 10 years ago, virtually nothing was known about black hole statistical mechanics. Today, we might suffer an embarrassment of riches [4]: we have many competing microscopic pictures, describing different states and different dynamics but all predicting the same thermodynamic behavior [5-13]. Nevertheless, some explanations are based on those theories that themselves remain to be experimentally verified, some resort to new hypotheses and do not start from first principles, and some are dependent on a choice of an adjustable parameter. In particular, some explanations for the Bekenstein–Hawking entropy cite the Hawking temperature formula and then actually argue in circle, because they are essentially related to the process of proof: taking the Hawking temperature $k_B T_H$ given by Eq. (1) as a function of the energy $E = Mc^2$, one can integrate the expression $dS_{BH} = dE/k_B T_H(E)$ to give Eq. (2). In other words, as far as the Bekenstein–Hawking entropy itself (rather than its quantum corrections) is concerned, those interpretations based on Eq. (1) cannot present more information than the seminal work of Bekenstein and Hawking, and then are not real interpretations.

In this work, a heuristic argument for the statistical origin of the Bekenstein-Hawking entropy is presented. A rigorous theoretical model will be presented in our next work. For simplicity, we confine ourselves to Schwarzschild black holes.

## II. SOME HEURISTIC CONSIDERATIONS

Now that the Bekenstein–Hawking entropy can be found within the traditional theory, its physical interpretation should hide in the traditional theory (from a logical point of view, see **Appendix A**). After all, historically, only when a new experimental result cannot be interpreted via the existing theory, a new hypothesis has to be introduced. Therefore, to seek for a convincing model for the statistical origin of the Bekenstein-Hawking entropy, the preferred way is to start from first principles and without resorting to any new hypothesis. That is, we should firstly attempt to deduce everything from known laws of physics, keep everything as simple as possible and only accept complications when they clearly become unavoidable. In the following, based on known laws of physics, we will present a heuristic argument for the statistical origin of the Bekenstein-Hawking entropy.

### A. General considerations

To study what the microscopic states responsible for the Bekenstein-Hawking entropy are, one should take the following basic facts into account:

1) Black holes have "no hair," no classical degrees of freedom that could account for their thermodynamic behaviors, which implies that the Bekenstein-Hawking entropy should be associated with those degrees of freedom excited by quantum effects;

2) Taking the Hawking temperature $k_B T_H$ as a function of the energy $E = Mc^2$, one can obtain the Bekenstein–Hawking entropy $S_{BH}$ by integrating $dS_{BH} = dE/k_B T_H(E)$, which implies that the most direct origin of the Bekenstein–Hawking entropy should be related to the Hawking radiation;

3) Seeing that the Bekenstein–Hawking entropy can be discovered by means of quantum field theory in curved spacetime, its statistical origin should lie in the first-order approximation of quantum gravity (i.e., quantum field theory in curved spacetime), only its quantum corrections require a detailed knowledge of quantum gravity. That is, at least for large black holes, it must be possible to explain black hole entropy without requiring the details of quantum gravity.

In view of the above reasons 1)-3), it is advisable to attribute the statistical origin of the Bekenstein–Hawking entropy to vacuum fluctuations continually occurring in the vicinity of the horizon of a black hole, where the microstates (counted by the Bekenstein–Hawking entropy) are related to all possible configurations of vacuum fluctuations. Vacuum fluctuations can result in (virtual) particle-antiparticle pairs, where each of pairs as an entire unit can be regarded as a boson. Because of conservation laws, such boson behaves as a scalar particle and a quantum entanglement pair. As we know, a dissipative environment or interference from measuring instruments can induce a mixed state at zero temperature, which is responsible for quantum decoherence and entanglement entropy. For our issue, the presence of the event horizon is responsible for black hole entropy: as for quantum entanglement pairs produced out of vacuum fluctuations of quantum fields present around a black hole, some particles in these pairs enter the black hole while their partners move to infinity, as measured by a static observer at infinity, the black hole behaves as emitting particles moving away from the horizon (i.e., the Hawking radiation), and possesses entanglement entropy. In our next work, we will present a rigorous theoretical model that shows that such entanglement entropy is exactly the Bekenstein–Hawking entropy, which in agreement with the above reason 2). In the model we will present a new treatment about the divergence of black hole entropy on the horizon.

In a word, based on quantum field theory in curved spacetime and without introducing any new hypothesis, it is inevitable for us to attribute the Bekenstein–Hawking entropy to entanglement entropy.

**B. Gravitational restriction on Heisenberg's uncertainty principle**

Heisenberg's uncertainty relations just provide a lower limit for the product between the uncertainties in (generalized) momentum and position. However, once the gravitational interaction is taken into account, one will find there is another restriction condition imposed on the uncertainties.

In Minkowski spacetime, Heisenberg's uncertainty principle tells us there is an uncertainty relation between momentum and position (or energy and time) as follows:

$$\Delta x \Delta p \geq \hbar/2, \quad \Delta t_\mathrm{D} \Delta E \geq \hbar/2, \qquad (3)$$

where $\Delta p$ and $\Delta x$ are respectively the uncertainties in momentum and position, $\Delta E$ and $\Delta t_\mathrm{D}$ are respectively the uncertainties in energy and time, and $t_\mathrm{D}$ denotes a *dynamical time*. The time-energy uncertainty relation has been a controversial issue and has many different versions, which is due to the nonexistence of a self-adjoint time operator [13, 14]. The Mandelstam–Tamm version of the time-energy uncertainty relation is the most widely accepted nowadays, where the time uncertainty $\Delta t_\mathrm{D}$ is given by a characteristic time associated with some dynamical variables (and then its physical origin is not unique). In general, let $f$ be a physical quantity whose operator ($\hat{f}$, say) does not depend explicitly on the parametric time $t$, using the Heisenberg equation of motion of $\hat{f}$, one can show that the uncertainty $\Delta f$ in $f$ and the uncertainty $\Delta E$ in energy satisfy the uncertainty relation:

$$\Delta f \Delta E \geq \hbar \left| < \mathrm{d}\hat{f}/\mathrm{d}t > \right|/2, \tag{4}$$

where $< \mathrm{d}\hat{f}/\mathrm{d}t >$ denotes the quantum mechanical average of $\mathrm{d}\hat{f}/\mathrm{d}t$. Define a characteristic time associated with the uncertainty $\Delta f$ as follows:

$$\Delta t_\mathrm{D} = \Delta f \Big/ \left| < \mathrm{d}\hat{f}/\mathrm{d}t > \right|. \tag{5}$$

Substituting Eq. (5) into Eq. (4), Eq. (4) will become the time-energy uncertainty relation. On the other hand, though Heisenberg's uncertainty relations are always explained in terms of a measurement process, they actually have many different interpretations and then have different physical origins. For example, the energy uncertainty $\Delta E$ may stem from energy fluctuations that can result in an energy level transition or particle decay etc., for the moment the $\Delta E$ has nothing to do with any measurement process.

The event horizon of a black hole is a boundary defined by lightlike geodesics. For our purpose, we will consider the case of $\hat{f} = x$ with $\left| < \mathrm{d}\hat{f}/\mathrm{d}t > \right| = c$, for the moment Eq. (5) shows that the time uncertainty $\Delta t_\mathrm{D}$ is related to the position uncertainty $\Delta x$ via $\Delta t_\mathrm{D} = \Delta x/c$, and Eq. (4) becomes $\Delta x \Delta E \geq \hbar c/2$, which implies that if an energy fluctuation with the magnitude of $\Delta E$ occurs within $\Delta x$, the condition of $\Delta x \Delta E \geq \hbar c/2$ must be true. Let us stress that, the position uncertainty $\Delta x$ around the mean position implies that the real position $x$ varies from its mean value in both directions.

According to quantum mechanics, the uncertainty relation $\Delta x \Delta E \geq \hbar c/2$, as an inequality, only its lower limit of $\hbar c/2$ is given, while both $\Delta x$ and $\Delta E$ have no upper limit. However, according to general relativity, within a given space interval of $\Delta x$, the energy fluctuation $\Delta E$ has an upper limit by forming a Schwarzschild black hole with the Schwarzschild radius of $\Delta x = 2G\Delta E/c^4$ and the mass of $\Delta E/c^2$, where the black hole takes the mean position of $\bar{x}$ as its center of sphere. That is, one has $\Delta E \leq \Delta x c^4/2G$.

Contrarily, once the energy fluctuation $\Delta E$ occurring within $\Delta x$ is larger than $\Delta x c^4/2G$, a black hole with the Schwarzschild radius of $\Delta x' = 2G\Delta E/c^4 > \Delta x$ would be formed (also taking $\bar{x}$ as its center of sphere). Because of the event horizon, the position uncertainty is no longer $\Delta x$, but rather $\Delta x' > \Delta x$, i.e., the $\Delta x$ is covered by the black hole. In terms of the present position uncertainty $\Delta x'$, the relation of $\Delta E \leq \Delta x' c^4/2G$ is still valid. Therefore, let the position uncertainty be denoted constantly by $\Delta x$, within which the energy fluctuation $\Delta E$ must satisfy the relation of $\Delta E \leq \Delta x c^4/2G$. From these discussions it follows that:

$$(\Delta x)^2 c^3/2G \geq \Delta x \Delta p \geq \hbar/2, \quad (\Delta t_D)^2 c^5/2G \geq \Delta t_D \Delta E \geq \hbar/2. \tag{6}$$

Eq. (6) can also be expressed as,

$$\begin{cases} \Delta x \Delta p \geq \hbar/2 \\ \Delta p/\Delta x \leq c^3/2G \end{cases}, \quad \begin{cases} \Delta t_D \Delta E \geq \hbar/2 \\ \Delta E/\Delta t_D \leq c^5/2G \end{cases}. \tag{7}$$

We call Eq. (7) *the complete uncertainty relations* (the complete uncertainty relations between other canonical conjugate pairs can be obtained in a similar way). That is, once quantum mechanics and general relativity are taken into account simultaneously, one should add another restriction to the traditional uncertainty relations.

Some interesting results can be derived from Eq. (6) or Eq. (7). For example, using Eq. (6) and let the upper limit be equal to the lower limit, one can obtain the Planck scales. Because the upper limit comes of gravity while the lower limit stems from quantum mechanics, the Planck scales stand for the confluence between gravity and quantum mechanics. On the other hand, it follows from Eq. (6) that $(\Delta x)^2 c^3/2G \geq \hbar/2$, which implies that the Planck length $l_p$ represents the least position uncertainty: $\Delta x \geq l_p = \sqrt{\hbar G/c^3}$. In fact, one can understand the Planck length from several different points of view: 1) the Planck length is the only length that can be formed from the constants $c$, $G$, and $\hbar$; 2) let the Compton wavelength of a particle be equal to the double Schwarzschild radius of the particle, one can obtain the Planck length; 3) the possibility of forming a black hole means that the uncertainty in position is bounded below by the Planck length, and then the Planck length represents the smallest length that can be operationally defined. It should be stressed that, Eq. (6) or (7) is obtained by starting from first principles and without resorting to any additional hypothesis, but it can achieve the same goal as which obtained by the so-called Generalized Uncertainty Principle [15-20].

Moreover, let us define

$$\Delta F \equiv \Delta p / \Delta t_\text{D} = \Delta E / \Delta x. \tag{8}$$

We regard $\Delta F$ as a force stemming from vacuum fluctuations. According to our discussions above, a vacuum fluctuation satisfying the condition of $\Delta F = c^4/2G$ will induce a black hole.

When we try to attribute the statistical origin of the Bekenstein–Hawking entropy to vacuum fluctuations occurring in the vicinity of the horizon of a black hole, Eq. (7) presents us with a natural high-energy cut-off.

**C. A heuristic consideration for the Bekenstein–Hawking entropy**

In the present work we just provide a heuristic consideration for the statistical origin of the Bekenstein–Hawking entropy, which is helpful for us to search for a rigorous model. It should be emphasized that the Hawking radiation can be viewed from several different perspectives being equivalent to one another, associated with which, in the following we discuss the Bekenstein–Hawking entropy in terms of vacuum fluctuations rather than entanglement entropy. According to the preceding considerations, the Hawking radiation gives a clue to the statistical origin of the Bekenstein-Hawking entropy, such that it is advisable to attribute the statistical origin of the Bekenstein–Hawking entropy to vacuum fluctuations continually occurring in the vicinity of the horizon of a black hole, where the microstates (counted by the Bekenstein–Hawking entropy) are related to all possible configurations of these vacuum fluctuations. Seeing that these vacuum fluctuations can result in (virtual) particle-antiparticle pairs and each of these pairs as an entire unit behaves as a scalar particle, we call them the embryos of particle, and all possible vacuum fluctuations forms the thermodynamic system of embryos (assume that it forms a grand canonical ensemble with the chemical potentials vanishing. In fact, the brick-wall model of 't Hooft is also based on a grand canonical ensemble with the chemical potentials vanishing [21]).

As a heuristic consideration, our treatment is presented as follows:

1) In contrast with the brick-wall model of 't Hooft, we will not resort to a grand canonical ensemble with null chemical potentials, but rather to a microcanonical ensemble with an extended phase space possessing the symplectic form of $dt \wedge dE$ (as a result, the extended phase space induced by all possible vacuum fluctuations is composed of the $n$-th power of $\Delta E \Delta t$, where $n=0,1,2\ldots$, represents the number of virtual bosons, $\Delta t$ is a characteristic time associated with a position uncertainty $\Delta x$ within which the energy fluctuation $\Delta E$ occurs);

2) The gravitational restriction on Heisenberg's uncertainty principle (given by Eq. (7)) will be taken into account, by which a natural ultraviolet cutoff is introduced;

3) We just consider all possible configurations of vacuum fluctuations, and do not concern the number of different field species which exist in Nature (such "species problem" and the UV divergent problem will be treated in our next work);

4) In contrast with the brick-wall model of 't Hooft, to avoid arguing in circle we try to get the Bekenstein–Hawking entropy without applying the Hawking temperature formula.

In our simplification, as a rough estimate, we regard a vacuum fluctuation near the event horizon as an energy fluctuation of $\Delta E$ occurring within the corresponding position fluctuation of Δx. In Sect. III, we will prove that the position fluctuations along the radial direction of a black hole approach to zero in the eyes of the observer stationary near the event horizon, which is due to the fact that, the coordinate transformation from a Kruskal-Szekeres reference frame to the stationary reference frame near the event horizon is equivalent to a squeezed transformation, and as viewed from the observer the surface gravity approaches to infinity. Therefore, only the position fluctuations along the event horizon are under consideration. Along the horizon, the largest uncertainty in position is $2\pi r_s$ ($r_s$ is the Schwarzschild radius), and then one has $0 \leq \Delta x \leq 2\pi r_s$. For the moment Eq. (7) implies that

$$\Delta E \Delta x \leq (\Delta x)^2 c^4 / 2G \leq (2\pi r_s)^2 c^4 / 2G. \tag{8}$$

Then vacuum fluctuations occurring at the horizon satisfy

$$(\Delta E \Delta t)_{max} = (2\pi r_s)^2 c^3 / 2G. \tag{9}$$

The extended phase space of single particles also consists of cells of magnitude of $2\pi\hbar$, the total number of the microstates contained in the extended phase space with the volume of $\Delta E \Delta t$ is

$$\Delta \Gamma_1 = \frac{1}{2\pi\hbar} (\Delta E \Delta t)_{max} = \frac{\pi r_s^2 c^3}{\hbar G} = \frac{4\pi M^2 G}{\hbar c}. \tag{10}$$

Here, as for a given black hole, all the vacuum fluctuations with $\Delta E > 2\pi r_s c^4 / 2G$ are not taken into account, because such vacuum fluctuations can create a new black hole with the radius larger than $(\Delta x)_{max} = 2\pi r_s$, and then the new black hole can swallow the whole horizon of the original black hole. Why it is $\Delta E > 2\pi r_s c^4 / 2G$ rather than $\Delta E > 2 r_s c^4 / 2G$? Firstly, the former is a natural result of Eq. (7); secondly, the gravitational field of the new black hole would deform the horizon of the original black hole, such that the case of $\Delta E > 2\pi r_s c^4 / 2G$ is more likely to result in a complete swallow for the original black hole.

Likewise, the microstates of *n* virtual bosons are described by the extended phase space formed by $(\Delta E \Delta t)^n$ ($n$=0, 1, 2…), where the extended phase space of *n* virtual bosons consist of cells of magnitude of $(2\pi\hbar)^n$, the total number of all possible microstates of *n* virtual bosons is

$$\Delta \Gamma_n = \frac{(\Delta \Gamma_1)^n}{n!}, \quad n = 0, 1, 2..., \tag{11}$$

where the result has been divided by the number of possible permutations of *n* virtual bosons (which is *n*!), which is due to the fact that the microstates occupying the same phase-space cell are indistinguishable, and then vacuum fluctuations occupying the same phase-space cell create identical bosons. Using Eqs. (10) and (11), the total number of the microscopic states of all possible vacuum fluctuations should be

$$\Gamma = \sum_{n=0}^{+\infty} \Delta \Gamma_n = \sum_{n=0}^{+\infty} \frac{(\Delta \Gamma_1)^n}{n!} = \exp(\Delta \Gamma_1). \tag{12}$$

Then the entropy induced by vacuum fluctuations near the horizon is

$$S_{BH} = \ln \Gamma = \Delta \Gamma_1 = A_h / 4 l_p^2. \tag{13}$$

Here the temperature is measured in energy units, such that the factor $k_B$ is not included in the definition of entropy. Eq. (13) is exactly the Bekenstein–Hawking entropy. In contrast with the brick-wall model of 't Hooft, to get the Bekenstein–Hawking entropy we do not apply the Hawking temperature formula and then avoid arguing in circle. Moreover, in our case, the cut-off is naturally provided by Eq. (7) (from a physical point of view it is the black holes that should provide for a natural cut-off all by themselves). In the brick-wall model, there is a cut-off of short distance from the horizon given by hand.

**D. Entropy, phase space and the Area Law of black hole entropy**

In classical statistical mechanics, a microstate corresponds to a certain point in phase space. In quantum mechanics, Heisenberg's uncertainty principle prevents the notion of a system being characterized by a point in phase space, and only domains of minimum area $2\pi\hbar$ in phase space are allowed. Quantum mechanics describes a microscopic system in terms of a state vector $|\psi\rangle$ or a density operator $\hat{\rho}$. However, there exists a representation of quantum mechanics which brings out directly the properties of a quantum state. This representation lives in phase space and rests on the concept of the phase space distributions (such as Wigner function, P-function, and Q-function, etc., all of them can be described by the off-diagonal or diagonal matrix elements of a density operator $\hat{\rho}$ in different bases) [22]. These phase space distributions allow us to calculate quantum mechanical expectation values using concepts of classical statistical mechanics (there is an infinite amount of phase space distribution functions that achieve this goal). As a result, both in classical statistical mechanics and quantum statistical mechanics, entropy lives in phase space essentially. BTW, in **Appendix B** we show that, probability as an observable, its quantum-mechanical counterpart (i.e., probability operator) is exactly the so-called density operator, such that a reduced density operator plays the role of a marginal probability operator.

For our purpose, let us recall some materials [23]. Let $M^n$ be the configuration space of a mechanical system; the corresponding manifold $M$ has local coordinates $q = (q^1, q^2, ..., q^n)$. The phase space is the cotangent bundle $T^*M$ with local coordinates $(q, p) = (q^1, q^2, ..., q^n, p_1, p_2, ..., p_n)$. Introduce the notation $x^i = q^i$, $x^{n+i} = p_i$, $i = 1, 2..., n$. On $T^*M$ we have the Poincaré 1-form $\omega^1 = p_i dq^i$ and the resulting 2-form $\omega^2 = dp_i \wedge dq^i$. A 2-form $\omega^2$ on an even dimensional manifold $M^{2n}$ is called symplectic provided it satisfies (1) $d\omega^2 = 0$ and (2) $\omega^2$ is nondegenerate (in local coordinates $x$, $\det(\omega_{ij}^2) \neq 0$). Every cotangent bundle is a symplectic manifold. The 2n-form $\omega^{2n} \equiv \pm n! dq^1 \wedge ... \wedge dq^n \wedge dp_1 \wedge ... \wedge dp_n = \omega^2 \wedge ... \wedge \omega^2$ is the Liouville or symplectic volume for the phase space. Since $\omega^2$ is a well-defined 2-form on any cotangent bundle, this 2$n$-form is actually independent of the local coordinates $q = (q^1, q^2, ..., q^n)$ used on $M^n$.

In particular, time-dependent Hamiltonians $H = H(q, p, t)$ is a function on the extended phase space $T^*M \times R$ with Poincaré 1-form $\Lambda^1 = p_i dq^i - H dt$ and the resulting 2-form $\Lambda^2 = dp_i \wedge dq^i - dH \wedge dt$. However, when the Hamiltonian of a thermodynamic system is time-dependent, by means of a canonical ensemble (or grand canonical ensemble) we can still work on the corresponding phase space, and do not resort to the extended phase space. A black hole should be considered as a grand canonical ensemble, because the number of particles and the energy of the system are not constants (but all chemical potentials may be kept close to zero, just as considered in the brick-wall model of 't Hooft).

Now let us come back to our topic. Usually, people are of opinion that the area proportionality of black hole entropy (the 'Area Law') differs from *the volume proportionality* of familiar thermodynamic systems, and regard it as an evidence that all the information contained in a volume can be completely carried by the boundary of the volume, such that some people interpret black hole entropy by introducing the hologram principle.

However, the entropy of a grand canonical ensemble is proportional to the measure of the corresponding phase space. Since black hole's radius is proportional to its mass, one has $S_{BH} \propto dr dp \propto A_h$, where d$r$ is along the radial direction of the black hole, both of the integrals respectively with respect to $r$ and $p$ are proportional to the mass of the black hole, and then the Area Law of black hole entropy is true (here we just give a heuristic statement, a rigorous argument will be presented later). That is, if the configuration-space measure (such as d$r$) is proportional to the momentum-space measure (such as d$p$), then the phase-space measure (such as d$r$d$p$) would be proportional to a configuration-space area. As a result, if black hole entropy obeyed the law of usual thermodynamic systems (as grand canonical ensembles), it would not obey the so-called volume proportionality, but rather be proportional to the cube of the horizon area, i.e., $S_{BH} \propto \mathrm{d}^3 x \mathrm{d}^3 p \propto A_h^3$. Contrarily, if we declare that the entropy $S$ of usual thermodynamic systems satisfies the volume law because of $S \propto \mathrm{d}^3 x \mathrm{d}^3 p \propto V \mathrm{d}^3 p$, then we should say that black hole entropy $S_{BH}$ satisfies *the line law* because of $S_{BH} \propto \mathrm{d}x \mathrm{d}p \propto L \mathrm{d}p$ (instead of the area law). Here, it is important that only on an equal footing can we reasonably compare black hole with usual thermodynamic systems, from which we can find the real distinction between them. That is, to compare black hole with usual thermodynamic systems, one should assume that the usual thermodynamic systems satisfy the same condition: their radius is proportional to their mass.

It follows from the above discussions that, the statement that "the area law of black hole entropy differs from the volume law of familiar thermodynamic systems" is not appropriate. Instead, one should say that "the area law of black hole entropy differs from the area-cube law of familiar thermodynamic systems", or equivalently, that "the line law of black hole entropy differs from the volume law of familiar thermodynamic systems", which no longer implies that all the information contained in a volume can be completely carried by the boundary of the volume. Even if for the usual thermodynamic system, the case that entropy is proportional to the spatial volume is just a particular case. In a word, it is not necessary to interpret black hole entropy by introducing the hologram principle.

## III. SUMMARY

To summarize, the Bekenstein–Hawking entropy can directly be derived from the Hawking temperature formula, which implies that its statistical origin is associated with the Hawking radiation and related to the gravitational field of black hole via Unruh effect, and then it is advisable to attribute the statistical origin to vacuum fluctuations in the vicinity of the horizon, where vacuum fluctuations create virtual particle-antiparticle pairs, and each of virtual pairs as an entire unit can be regarded as a virtual boson (because of conservation laws, such boson should be a virtual scalar particle). These virtual bosons may be off-shell and do not obey the equation of motion of quantum fields, we account their microstates via the volume of (extended) phase space occupied by the virtual bosons, which is different from the brick-wall model based on the equation of motion of a scalar field in the Schwarzschild spacetime. In particular, in contrast with the brick-wall model of 't Hooft, we get the Bekenstein–Hawking entropy without applying the Hawking temperature formula (such that we avoid arguing in circle), by which the statistical origin of the Bekenstein–Hawking entropy obtains a genuine interpretation.

In our next work, we will turn to the entanglement-entropy interpretation of black hole entropy, we may emphasize the following fact that: The phase-space volume occupied by a given set of $N$ identical particles at a given event in spacetime is independent of the local Lorentz frame in which it is measured. Moreover, as the same $N$ particles move through spacetime along their geodesic world lines (and through momentum space), the volume they span in phase space remains constant. The phase-space volume occupied by a given swarm of $N$ particles is independent of location along the world line of the swarm ("Liouville's theorem in curved spacetime") [24]. In particular, some questions presented in Ref. [25] will be taken into account in our next work.

**ACKNOWLEDGMENTS**

The author would like to thank Professor J. D. Bekenstein and Professor S. Carlip for their helpful comments and valuable suggestions. This work was supported by the Fundamental Research Funds for the Central Universities (Grant No: ZYGX2010X013).

*E-mail:   zywang@uestc.edu.cn

## APPENDIX A

### Principle of interpretation

Assume that a theory is equivalent to an axiomatics consisting of *N* axioms, $A_1, A_2, ..., A_N$, for the moment the theory can be denoted as the set of $\{A_1, A_2, ..., A_N\}$, all theorems or true propositions of the theory can be derived from some of the *N* axioms. Let $E(A_i)$ ($i = 1, 2, ..., N$) represent the *interpretation* of the axiom $A_i$, which can be regarded as a set (an axiom may have many different but equivalent interpretations, we take them as the same interpretation). Usually, an axiom is abstract, while its interpretation can be described via its concrete representations or physical realizations. Now, assume that there is a true proposition *P* in the axiomatics, which implies that it can be derived from *M* axioms, say

$$B_1, B_2, ... B_M \mapsto P, \ 1 \leq M \leq N. \tag{a1}$$

The set of $\{B_1, B_2, ... B_M\}$ formed by the *M* axioms should satisfy

$$\{B_1, B_2, ... B_M\} \subset \{A_1, A_2, ..., A_N\}. \tag{a2}$$

Let $E(P)$ denote the interpretation of the proposition *P*, it should satisfy the relation:

$$E(P) = \bigcup_{j=1}^{M} E(B_j) \subset \bigcup_{i=1}^{N} E(A_i), \tag{a3}$$

where $\bigcup_{j=1}^{M} E(B_j)$ denotes the union set of $E(B_j)$ ($j = 1, 2, ..., M$), and so on. We call Eq. (a3) the *principle of interpretation*.

## APPENDIX B

### Density operator as a probability operator

Though probability has been introduced mathematically, it is actually a physical quantity (an observable) related to the concept of information or entropy. In particular, people have recently interpreted gravity as an entropic force, which further endows probability with a dynamical meaning. However, in quantum mechanics, observables are represented by linear self-adjoint operators acting on a Hilbert space of quantum states, which implies that a quantum-mechanical counterpart of probability, say, a self-adjoint probability operator, should exist. Then, how to assign an operator to the observable of probability in quantum mechanics?

To assign an operator to the probability of finding a system in a certain microstate, we introduce a self-adjoint probability operator $\hat{P}$ whose eigenvalues are nonnegative and whose trace (the sum of its eigenvalues) is one, that is

$$\hat{P}|\varphi_n\rangle = P_n|\varphi_n\rangle, \quad \hat{P} = \hat{P}^\dagger, \quad \sum_n P_n = 1, \quad 0 \leq P_n \leq 1, \tag{b1}$$

where $n = 1, 2...$, the eigenvalue $P_n$ represents the probability that the system assumes the corresponding eigenstate $|\varphi_n\rangle$ ($n = 1, 2...$), and corresponds to a possible measurable value of the observable probability. As the eigenvectors of the self-adjoint operator $\hat{P}$, $|\varphi_n\rangle$'s satisfy the following orthonormality and completeness relations (let $I$ denote an unit operator)

$$\langle\varphi_m|\varphi_n\rangle = \delta_{mn}, \quad \sum_n |\varphi_n\rangle\langle\varphi_n| = I, \quad m, n = 1, 2, .... \tag{b2}$$

Then the set $\{|\varphi_n\rangle\}$ forms a complete orthonormal basis that spans the Hilbert space of the considered system. Using Eqs. (b1) and (b2) one can show that the quantum-mechanical average of $\hat{P}$ in a pure state of $|\Psi\rangle$ reads

$$\bar{P} = \langle\Psi|\hat{P}|\Psi\rangle = \sum_n P_n|\lambda_n|^2 = \sum_n P_n\langle\varphi_n|\hat{P}_\Psi|\varphi_n\rangle, \tag{b3}$$

where $|\lambda_n|^2 = |\langle\Psi|\varphi_n\rangle|^2$ represents the probability of obtaining the state $|\Psi\rangle$ from $|\varphi_n\rangle$,

$\hat{P}_\Psi = |\Psi\rangle\langle\Psi|$ is a projection operator. Eq. (b3) shows us that, as the quantum-mechanical average of $\hat{P}$ in the state of $|\Psi\rangle$, $\bar{P} = \langle\Psi|\hat{P}|\Psi\rangle$ is equivalent to the statistical average of $\langle\varphi_n|\hat{P}_\Psi|\varphi_n\rangle$ with respect to the probability of $P_n$, where $\langle\varphi_n|\hat{P}_\Psi|\varphi_n\rangle$ is the quantum-mechanical average of $\hat{P}_\Psi = |\Psi\rangle\langle\Psi|$ in the state of $|\varphi_n\rangle$. Therefore, Eq. (b3) concerns two kinds of average simultaneously. From another point of view, a physical quantity's average depends on its probability distribution, which is also valid for a probability itself, and then one can talk about a probability distribution of a probability. In quantum mechanics, the observed value of a physical quantity can be obtained by averaging over all possible quantum states, which is also valid for a probability itself. However, in contrast to other physical quantities, the average probability $\bar{P}$ in Eq. (b3) represents not only the average of the probability $|\lambda_n|^2 = |\langle\Psi|\varphi_n\rangle|^2$ with respect to the probability $P_n$, but also the average of the probability $P_n$ with respect to the probability $|\lambda_n|^2 = |\langle\Psi|\varphi_n\rangle|^2$. Nevertheless, according to probability theory, the average of a probability is the probability itself, which is related to the multiplication theorem of probability.

It follows from Eq. (b3) that the probability operator $\hat{P}$ has the *spectral representation* (or *spectral decomposition*):

$$\hat{P} = \sum_n P_n |\varphi_n\rangle\langle\varphi_n|. \tag{b4}$$

Eq. (b4) shows that the operator property of $\hat{P}$ is carried by $\hat{\mu}(n) = |\varphi_n\rangle\langle\varphi_n|$ (i.e., the *spectral measure* of $\hat{P}$). Let $\{|\psi_m\rangle\}$ represent another complete orthonormal basis of the Hilbert space, using $\sum_m |\psi_m\rangle\langle\psi_m| = I$, $\sum_n P_n = 1$ and $\langle\varphi_m|\varphi_n\rangle = \delta_{mn}$ one has

$$\text{Tr}(\hat{P}) = \sum_m \langle\psi_m|\hat{P}|\psi_m\rangle = 1. \tag{b5}$$

According to quantum statistics, in an ensemble formed by $\{|\varphi_m\rangle, P_m\}$, the quantum statistical average of a self-adjoint operator $\hat{F}$ is defined as

$$\bar{F} = \sum_m P_m \langle \varphi_m | \hat{F} | \varphi_m \rangle. \tag{b6}$$

Using Eqs. (b2) and (b4) one can obtain

$$\bar{F} = \sum_m \langle \varphi_m | \hat{P}\hat{F} | \varphi_m \rangle = \mathrm{Tr}(\hat{P}\hat{F}). \tag{b7}$$

Eqs. (b4), (b5) and (b7) together show that the probability operator $\hat{P}$ is exactly a density operator. If $0 \leq P_n < 1$ for all $n = 1, 2...$, the set of $\{|\varphi_n\rangle, P_n\}$ forms a mixed ensemble; while, if $P_n = 1$ for one index $n$ and zero otherwise, one has a pure ensemble that assumes the unique state $|\varphi_n\rangle$, and $\hat{P} = |\varphi_n\rangle\langle\varphi_n|$ becomes a projection operator. When the sum indices in Eq. (b4) are continually varying, the summation becomes an integral and $P_n$ is replaced by a probability density.

Therefore, in our formalism, density operators are taken as quantum-mechanical counterparts of classical probabilities (i.e., self-adjoint probability operators). In particular, as observables, their quantum statistical averages can also be obtained by Eq. (b7). For example, in an ensemble formed by $\{|\varphi_m\rangle, P_m\}$, the projection operator $\hat{P}_\Psi = |\Psi\rangle\langle\Psi|$ as a probability operator, its average is given by Eq. (b7)

$$\bar{P}_\Psi = \mathrm{Tr}(\hat{P}\hat{P}_\Psi) = \sum_n P_n \langle \varphi_n | \hat{P}_\Psi | \varphi_n \rangle. \tag{b8}$$

Then one has $\bar{P}_\Psi = \bar{P}$ (see Eq. (b3)), which further implies that Eq. (b3) not only represents the average of the probability $|\lambda_n|^2 = |\langle\Psi|\varphi_n\rangle|^2$ with respect to the probability $P_n$, but also the average of the probability $P_n$ with respect to the probability $|\lambda_n|^2 = |\langle\Psi|\varphi_n\rangle|^2$, just as mentioned before.

In the traditional terminology, a density operator is also called a state. However, for example, the density operator $\hat{P}_\Psi = |\Psi\rangle\langle\Psi|$ is an observable while the corresponding state

$|\varPsi\rangle$ not.

As we know, a quantum-mechanical operator possesses different matrix representations in the different bases of the same Hilbert space. Likewise, a probability operator (or density operator) can correspond to different probability distributions in different bases. For example, the Wigner function is related to the off-diagonal matrix elements of density operators (and then it is not positive definite and just a quasi-probability distribution), while some other phase space distributions (such as the Q-function are related to the diagonal matrix elements of a probability operator and then are positive definite. Mathematically, a classical probability can be described as a measure. Correspondingly, a probability operator can be described as a quantum measure operator via Eq. (4), its diagonal elements in an orthonormal basis form a probability distribution when this basis is used as the quantum sample space. In general, a density operator $\hat{\rho}$ plays the role of a probability operator $\hat{P}$: $\hat{\rho} = \hat{P}$, and a probability measure $\varphi(\hat{E}) = \text{Tr}(\hat{\rho}\hat{E})$ corresponds to the average of the operator $\hat{E}$, where $\hat{E}$ is always an especial probability operator, such that the probability measure $\varphi(\hat{E}) = \text{Tr}(\hat{\rho}\hat{E})$ can also be regarded as the average of the probability operator $\hat{\rho} = \hat{P}$. In classical probability theory, the marginal distribution of a subset of a collection of random variables is the probability distribution of the variables contained in the subset. Likewise, in the case of a composite quantum system consisting of two or more subsystems, by means of a reduced density operator one can construct a quantum description of just one of these subsystems, while ignoring the other subsystem(s). Based on the fact that a density operator is actually a probability operator (i.e., quantum-mechanical counterpart of probability), one can easily show that a reduced density operator plays the role of a marginal probability operator.